\documentclass[12pt,twoside,a4paper]{article}

\renewcommand{\d}{ \, {\rm d}}
\newcommand{\na}{ {\rm n.a.} }

\newcommand{\gu}{ {\rm g,u} }
\newcommand\lesssim{\mathrel{\hbox{\rlap{\hbox{\lower4pt \hbox{ $\sim$}}} \hbox{$<$}}}}

\usepackage{graphicx}

\begin{document}

\title{Improved LeRoy-Bernstein near-dissociation expansion formula. Tutorial application to photoassociation spectroscopy of long-range states}

\author{Daniel Comparat \\ Laboratoire Aim\'{e} Cotton,$^{\dag }$ CNRS II, B\^{a}t. 505, Campus\\
d'Orsay, 91405 Orsay cedex, France}
 
\date{\today}

\maketitle

\abstract{NDE (Near-dissociation expansion) including LeRoy-Bernstein formulas 
are improved by taking into account the multipole expansion coefficients and the non asymptotic 
part of the potential curve. 
Applying these new simple analytical formulas 
 to photoassociation spectra of cold alkali atoms, we improve the 
determination of the asymptotic coefficient, reaching a $1\%$ accuracy, for long-range relativistic  potential curve of diatomic molecules.}

\setcounter{page}{1} 

\section{Introduction}

The interaction
between two distant atoms has been first studied by Van der Waals and
London (for review see~\cite{Margenau39}).  This topic is often
discussed as a limiting case between the Hund  case (a) and (c)~\cite{Mulliken40,Chang67}.
  The study of such excitation
transfers~\cite{Dashevskaya69,Movre77} are related to long-range molecular states, 
 where the electronical potential $V(R)$ is fully
described by the asymptotic coefficients\footnote{LeRoy and Bernstein and some other authors
 use
 $V(R) \approx D- \sum_k {C_k \over R^k}$.}
\begin{equation}
V(R) \approx D+ \sum_k {C_k \over R^k}
\label{asym_form}
\end{equation}
 for sufficiently large internuclear distance $R$ \cite{cm:stwalley78b}.
Among
these long-range states Stwalley {\it et al.}~\cite{Stwalley78} discovered very
particular molecular states:  the so-called ''pure long-range state'', where both classical turning points are in this
asymptotic area. Great efforts have been devoted to the 
precise calculations of the asymptotic coefficients $C_k$~\cite{Chang67,Dalgarno66,Bussery85}.

Semiclassical formulas in diatomic molecular spectroscopy are powerful tools
(for a brief review see~\cite{LeRoy80b,Vigue82}).
Several molecular properties as rotational or
 vibrational progression and
kinetic energy are strongly determinate by the leading terms in  
equation~(\ref{asym_form}):
\begin{equation}
V(R) \approx D+ {C_n \over R^n}+{C_m \over R^m}+ \ldots 
\label{asy_form}
\label{Dv_mul}
\end{equation}
where $m>n$. In the course of this, article we shall suppose $n>2$.  In 1970, LeRoy and
Bernstein~\cite{LeRoy70} pioneer work, based on the Bohr quantization formula, made possible  to extract the leading coefficient $C_n$ from experimental vibrational progression. 
The LeRoy-Bernstein formula links 
the energy $E$ of the vibrational quantum number $v$
with the asymptotic behavior $D + C_n/R^n$ of the potential curve :
\begin{equation}
E = D - \left[
(v_D - v) \left(
 \sqrt{ \pi \over 2 \mu } 
 {  \Gamma( 1 + 1/n)
\over \Gamma(1/2 + 1/n )
} { \hbar   (n-2) \over (-C_n)^{1/n} }
\right)
\right]^{{2n \over n-2}}
\label{leroybern}
\label{lRbernst}
\end{equation}
where $\mu$ is the reduced mass of the system and $v_D$ is the
non-integer value of $v$
  at the dissociation energy $D$.
This kind of
near-dissociation expansion (NDE) semi-classical formula was 
extended to the rotational progression~\cite{LeRoy72} and
 the kinetic energy~\cite{Stwalley73}.
 The technique was also improved to include other 
coefficients in the asymptotic development~\cite{LeRoy80} and a quasi-complete
NDE theory was established~\cite{LeRoy94}. 
 The subject is still in progress: links
 with Quantum Defect Theory, scaling law for the density probability of presence of the vibrational wavefunction \cite{Vigue82,cm:masnou01}, and
 extension to two coupled channels and Lu-Fano plots have been
successfully investigated~\cite{Ostrovsky01,cm:kokoouline02}. The goal of this article is to improve part of the NDE theory.

Experimental studies of the long-range states \cite{cm:allard02}
have recently been renewed by the photoassociation (PA) spectroscopy of trapped cold atomic samples.
Trapping and cooling
of neutral atomic samples, based on radiation pressure, are well
established~\cite{VanderStraten99} techniques that led to further spectroscopic developments.
For instance, in 1987, Thorsheim {\em et al.}~\cite{Thorsheim87} proposed a new spectroscopic
technique: the photoassociation process where
 a pair of free cold atoms absorbs resonantly one photon
and produces an excited molecule in a well-defined ro-vibrational level.
The first experiments were realized in 1993 in sodium and rubidium. Since these pioneer works
all the alkali atoms (Li, Na, K, Rb and Cs) (for a review 
see~\cite{Weiner99,Stwalley99}) 
then hydrogen~\cite{Mosk99}, metastable helium~\cite{cm:herschbach00}, calcium~\cite{cm:zinner00} and ytterbium~\cite{cm:takasu02}
have been photoassociated.
 Preliminary results for heteronuclear alkali systems have also
been reported \cite{cm:shaffer99,cm:schloder01}.

 In a
dilute medium, as the one present in the magneto-optical trap,
the probability to find two atoms at
a distance $R$ is proportional to $4 \pi R^{2} e^{-V(R)/k_B T}$. Consequently as PA is a collisional process it is efficient only at
large interatomic distance $R$. Therefore PA is particularly well adapted to the study of long-range
molecular states. 

Because of the extremely narrow continuum energy distribution (on the order of $k_B T$),
the photoassociation free-bound transition between the two free cold atoms
($T \approx 100\, \mu$K) and 
the ro-vibrational excited states is resolved at the MHz range
($k_B T \approx h \times 2\,$MHz).
This leads to an extremely precise spectroscopy~\cite{Stwalley99}.
The kHz range has been achieved in rubidium
starting with an atomic Bose-Einstein condensate \cite{Wynar00}.
New available precise data from PA spectroscopy have stimulated the 
theoretical determination of more precisely values for the asymptotic 
coefficients~\cite{Marinescu94,Marinescu95,Marinescu96}. 

We shall present here new useful simple analytical
formulas to extract the leading coefficient $C_n$ of the multipolar expansion within a $1\%$ precision.
To illustrate the importance of such a calculation, let us  mention
that this term occurs in the expression of atomic lifetime $\tau$
of the first excited $p$ level of a dialkaly molecule :
 \begin{equation}
\tau = {3 \hbar c^3 \over
4 |C_3| \omega_{\rm at}^3}
\label{ato_lifetime}
\end{equation}
where $\hbar \omega_{\rm at}$ is the energy difference
 between the $p $ excited atomic state and the $s$ ground state.
Indeed, a precise $\tau$ value was obtained using a pure long-range expansion of the  $0_g^- (s+p_{3/2})$ potential curve 
 of dialkalis~\cite{Alexander96,Jones96,Wang97b,cm:dulieu02}.

This article is organized as follows.
Section~\ref{sec:LRB} is devoted to the fully detailed derivation of a first improved LeRoy-Bernstein  formula
 using three new estimations respectively for 
the asymptotic part, for the repulsive branch part and for the intermediate part of the potential curve $V(R)$.
In  section~\ref{other_mult} we take into account
the next multipole coefficient $C_m$.
We finally obtain our main results, the general formula~(\ref{final_formula_ter}) for all the semiclassical NDE expressions and the improved LeRoy-Bernstein formulas~(\ref{fit_LRB}) and (\ref{LRB_final}). We apply these results in section~\ref{sec:test} on the $0_g^- (6s+6p_{3/2})$ state of the cesium dimer (where $n=3$ and $m=6$). We will discuss in great detail in the appendix \ref{long_range} how to derive formula~(\ref{Dv_mul}) for all cases, so that this theory can easily be extended to other long-range states. Indeed one goal of this article is to give a self sufficient theoretical background helping people interested in using our new simple analytical NDE formula in the
interpretation of photoassociation data.  

\section{Improved LeRoy-Bernstein theory}
\label{sec:LRB}

One of the simplest way to assign a given spectrum with a molecular potential curve is to
isolate its vibrational progression and to extract an
experimental $C_n$ coefficient, and then compare it to the theoretical $C_n$ coefficient.
This popular method makes use of
the analytical semi-classical
LeRoy-Bernstein 
formula~(\ref{lRbernst}), that  we propose here to improve.
 
\subsection{BKW assumption}

We use the Jeffreys, Brillouin, Kramers
 and Wentzel ((J.)B.K.W)
semi-classical method  and
the Bohr
quantization condition 
 (e.g. see~\cite{LandauMQ}) for the vibrational level $v$ at energy $E=E_v$ of a reduced
 mass $\mu$ particle moving in a potential $V(R)$:
\begin{equation}
   v + {1 \over 2} = { \sqrt{2\mu} \over \pi \hbar } \int_{R_-}^{R_+}
 \sqrt{ (E-V(R))}  \d R
  \label{cond_quant_BKW} 
\end{equation}
 $R_-(E_v)$ and $R_+(E_v)$ are respectively the inner and outer classical turning point
 of the vibrational motion ($V(R_-) = V(R_+) = E$).
At the dissociation limit
 $E=D$ the non-integer vibrational number $v$ results of the formula
is noted $v_D$.

For
levels very close to the dissociation limit, the quantization condition is still a controversial subject~\cite{Gao99,Eltschka01,Boisseau01,Moritz01}.
For instance, it has been shown~\cite{Flambaum93} that
Bohr  quantification condition  should
be modify at the dissociation limit by adding a $1\over 2(n-2)$ term at
the $v+1/2$ one. But the
modification is of noticeable importance only
for the few last levels (typically within less than $10\,$GHz energy range from the dissociation limit)
of the potential~\cite{Boisseau98}, where relativistic retardation effects or
hyperfine structure appear, and where it is no more realistic to use the
LeRoy Bernstein  formula. Nevertheless, if needed we can furthermore improve the formula by taking into account this $1\over 2(n-2)$ term or by using
the third order semi-classical theory \cite{LeRoy80b} and  adding a 
$
{ \hbar \over 48 \pi  \sqrt{2 \mu} } \int_{R_-}^{R_+}
 { \frac{\partial^2 V(R)}{\partial R^2} \over (E-V(R))^{3/2} }  \d R
$
term in the quantization formula.

However, as reported in~\cite{Vigue82}, the relative
BKW accuracy is on the order of 
$$|E_v^{\rm BKW} - E_v^{\rm real} |/ E_v^{\rm real} \lesssim {1\over \pi^2 v^2}.$$
 Thus,
 for levels close to the dissociation limit where typically $v=100$,  
there is no need to improve the usual Bohr quantization condition to reach the $1\%$ accuracy
we are looking for. Consequently, 
 in the following, we  shall use the usual Bohr  quantification condition (\ref{cond_quant_BKW}) and we shall see
that other assumptions are less accurate than this one. 

\subsection{Role of the non asymptotic part}

\begin{figure}[ht]
\begin{center}
\resizebox{\textwidth}{0.5\textwidth}{
\includegraphics*[1.9cm,0.6cm][26.2cm,21.1cm]{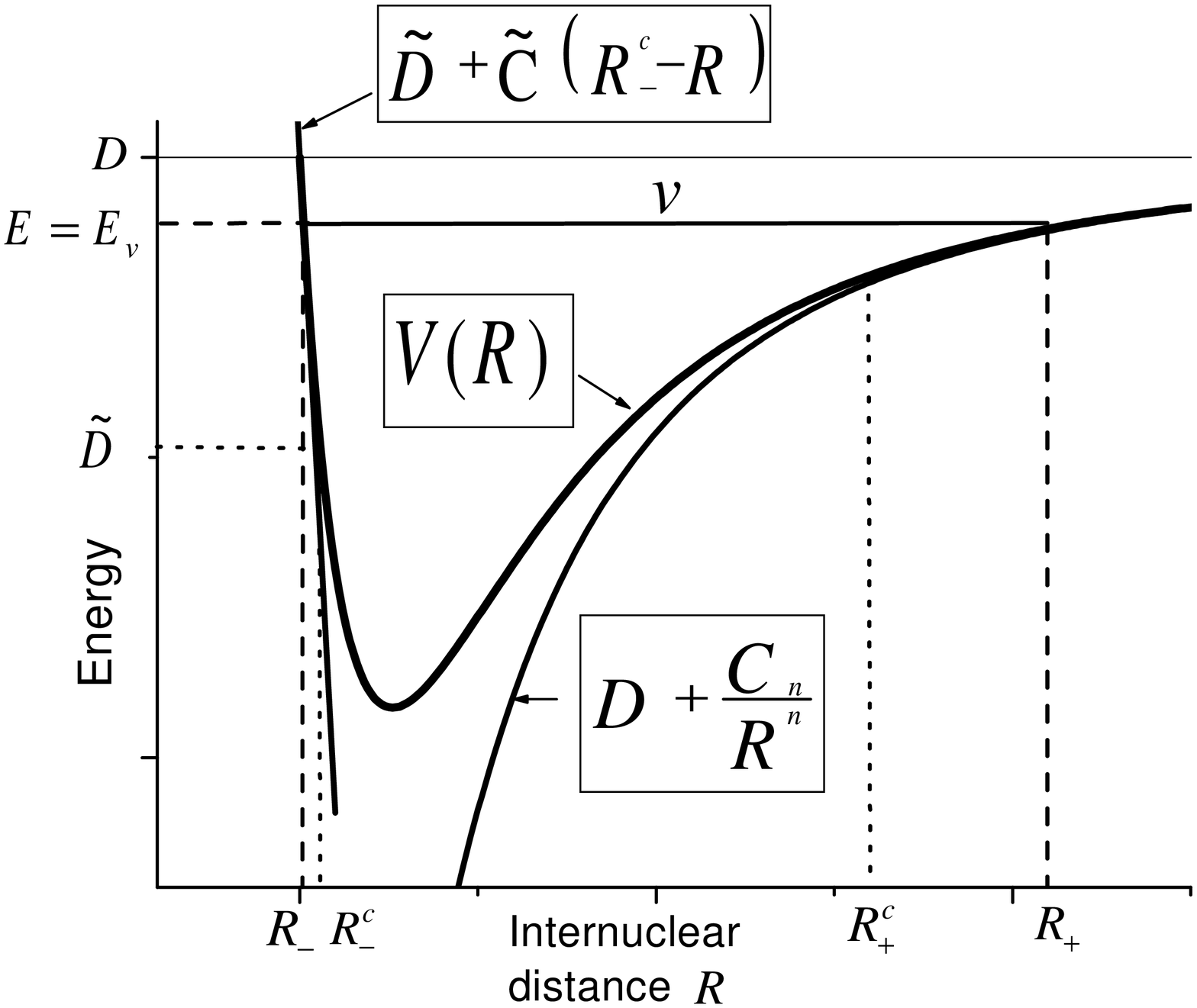} } 
\end{center}
\caption{The potential $V(R)$
 and the asymptotic potential
$D + C_n/R^n$.  Both potentials coincide
when $R$ is larger than the "cut-off"
$R_+^c$. 
The repulsive branch is model by a linear formula for $R \leq R_-^c$.}
\label{fig_Leroy}
\end{figure}

We define a ''cut-off'' outer-turning point $R_+^c$ where the potential $V$ can be written as:
\begin{equation}
V(R)  \stackrel{ \scriptstyle R>R_+^c }{\approx} D+{C_n \over R^n}
\label{asy_exp}
\end{equation}
within a given precision. The potential $V(R)$ and its asymptotic limit are represented in
 figure \ref{fig_Leroy}.
Our goal is to reach a $1\%$ precision then, if needed,
$R_+^c$ could be defined as:
\begin{equation}
0.01 {|C_n| \over (R_+^c)^n}  =  {|C_m| \over (R_+^c)^m}
\label{cut_off_def}
\end{equation}
 With typical values
as $n=3, m=6$, $|C_3| = 10 \, e^2 a_0^2$ and $|C_6|=10 000 \, e^2 a_0^5$, we obtain
$R_+^c \approx 45\,a_0$ (where $\,a_0 \approx 5.29\times10^{-11}\,$m and $e^2 =q_e^2 / (4 \pi \varepsilon_0)$).

It is now possible to separate the non-asymptotic part from
the asymptotic part ($R>R_+^c$). 
Taking the derivative of expression (\ref{cond_quant_BKW}), we obtain
(we use $x=R/R_+$):
\begin{eqnarray}
{\d v \over \d E} & = &
 { \sqrt{2 \mu } \over 2 \pi \hbar } 
{ (-C_n)^{1/n} \over (D - E)^{ {n+2 \over 2 n} } }
\int_{R_+^c/R_+ }^1  { x^{-n/2} \over \sqrt{1- x^{-n} } } 
   \d x
+ \nonumber \\
 & & \underbrace{
{ \sqrt{2 \mu } \over 2 \pi \hbar } 
\int_{R_-  }^{R_+^c }
 {1 \over \sqrt{E - V(R)} }
 \d R  
}_{ \left(
{\d v \over \d E} 
\right)^\na} =\frac{1}{\hbar \omega}
\label{asy_non_asym} \label{LRBERN}
\end{eqnarray}
where the subscript $\na$ is for non asymptotic and $\omega$ is the classical pulsation of the vibrational motion.

A physical insight on the role of the non asymptotic part can be obtained considering
the classical definitions of the velocity ${\rm v}$ and the impulsion $p$: 
$$
{\rm v}  = {p \over \mu} =  {\sqrt{2\mu(E - V(R))} \over \mu}.
$$
Equation~(\ref{LRBERN}) can then be  written as ${\d v \over \d E}  \propto \int_{R_- }^{R_+ }
 {1 \over \rm{v} }$. We can see in figure \ref{fig_Leroy} that the motion 
time is largely dominated by the asymptotic part of the potential, given by 
a multipole expansion as in formula (\ref{asy_exp}).
This classical discussion tells us that $\left(
{\d v \over \d E} \right)^\na \ll {\d v \over \d E}$ for levels close to the dissociation limit.
The next step is then to restrict ourselves to the levels close to the dissociation limit

\subsection{Near the dissociation limit}

\subsubsection{Correction in the asymptotic part}

For levels close to the dissociation limit we have the following inegality: $y= R_+^c/R_+ \ll 1$. We can then write:
\begin{eqnarray}
\int_{ y }^{1 } { x^{n/2} \over \sqrt{1- x^n } } 
   \d x
& = &
\int_0^1 { x^{n/2} \over \sqrt{1- x^n } } 
   \d x - \int^y_0 
{x^{n/2} \over  \sqrt{1-x^n   } } 
   \d x  \label{first_improv} \label{form_inte} \label{trois_equa}\\
&  = & { \sqrt{\pi} \over n} { \Gamma(1/2 + 1/n ) \over
 \Gamma( 1 + 1/n) } - { 2 \over n +2 } y^{{n+2
  \over 2} } 
(1+ {n+2 \over 6n+4} y^n + \ldots )   
\nonumber  \\
&\approx& { \sqrt{\pi} \over n} { \Gamma(1/2 + 1/n ) \over
 \Gamma( 1 + 1/n) } 
- 
 { 2 \over n +2 }  (R_+^c)^{ {n+2
\over 2} }  \left( {-C_n \over D-E} \right)^{- {n+2
\over 2n} }
\nonumber
\end{eqnarray}
where the main term (the only one taken into account in the ``usual'' LeRoy Bernstein law derivation)
and the first correction term has been kept. This correction term brings a real 
improvement. Indeed, if in the last formula
only the main term is kept, 
we need to take 
$ R_+ = 5 R_+^c $ ($y=1/5$) to obtain the integral (\ref{trois_equa}) value
at a $1\%$ level. Concequently, with $R_+^c \approx 45\,a_0$, reaching  a $1\%$ accuracy level with the 
the ''usual'' LeRoy-Bernstein formula requires to use levels with $R_+ > 200\,a_0$ where the Bohr quantization problems occur. On the contrary,
when using both terms, taking $ R_+ = 1.6  R_+^c$ 
is enough to reach the same precision level.
\label{pb_asym}

\subsubsection{Repulsive branch}
\label{non_asym_part}

To express 
 the non-asymptotic part in formula~(\ref{asy_non_asym}), we will model the inner wall by a linear function using another
cut-off $R_-^c$ (as indicated in  figure~\ref{fig_Leroy}) and two parameters 
$\tilde D $ and $ \tilde C$: 
\begin{equation}
V(R)  \stackrel{ \scriptstyle R<R_-^c}{\approx}  \tilde D + \tilde C(R_-^c - R)
\label{rep_lin}
\end{equation}
Other models (e.g.  
a potential with a $1/R$ behavior) do also lead to analytical 
formulas. 

The non-asymptotic integral in formula~(\ref{asy_non_asym}) can then be splitted in two integrals,
using $\int_{R_-  }^{R_-^c } = \int_{R_-  }^{R_-^c } + \int_{R_-^c}^{R_+^c }$.
The first integral is computed  analytically using 
formula~(\ref{rep_lin}):
\begin{equation}
\int_{R_- }^{R_-^c }
 {1 \over \sqrt{E - V(R)} }
 \d R  = { 2 \sqrt{E - \tilde D}  \over  \tilde C} \approx
 { 2 \sqrt{D - \tilde D}  \over  \tilde C}
\label{repu_branch}
\end{equation}
We have moreover use the approximation $E - \tilde D \approx D - \tilde D$, because 
we are dealing with levels close to the dissociation limit (see figure \ref{fig_Leroy}).
Better accuracy could be achieved by keeping $E$ in expression (\ref{repu_branch}).

\subsubsection{Intermediate region}

 For the second integral $\int_{R_-^c}^{R_+^c }$, 
we use the assumption $E - V(R) \approx D -V(R)$ that is valid in the intermediate
region $R_-^c < R <R_+^c$ (see figure~\ref{fig_Leroy}). Thus, this second integral
becomes simply a number and does not vary with $E$. If needed, it can be computed using for instance
 a model potential like Morse, Lennard-Jones or a quadratic one. 
Finally the non asymptotic part becomes:
\begin{eqnarray} 
 \left(
{\d v \over \d E} \right)^\na
 &\approx & 
 { \sqrt{2 \mu } \over  2 \pi \hbar } \bigg[
{ 2 \sqrt{D - \tilde D}  \over  \tilde C} 
\nonumber \\
& & +   
\int_{R_-^c }^{R_+^c  }
{ 1 \over \sqrt{ D - V(R)} }
 \d R \bigg]
\label{non_asy_eq}
\end{eqnarray}

\subsection{Improved LeRoy-Bernstein formula}

Using formula~(\ref{trois_equa}) in equation~(\ref{asy_non_asym}) and integrating it using 
 expression~(\ref{non_asy_eq}) leads to the improved formula:
\begin{equation}
v_D - v  \approx 
 H_n^{-1}  (D - E )^{{n-2 \over 2n}}  + \gamma ( D -  E )
\label{lRbernst_mod} 
\end{equation}
where 
\begin{equation}
 H_n^{-1}  = \sqrt{2 \mu \over \pi } {
 (-C_n)^{1/n} 
\over  \hbar (n - 2) } 
 { \Gamma({n+2 \over 2n} ) \over \Gamma( {n+1 \over n} )
}
\label{def_Hn}
\end{equation}
and $\gamma$ is an extra parameter strongly related to $R_+^c$ defined as follows:
\begin{eqnarray} 
\gamma & = & - { \sqrt{2 \mu } \over ( n +2 )  \pi \hbar } (- C_n)^{-1/2 }  (R_+^c)^{ {n+2 \over 2} } +  \label{gamma_form} \\
& & { \sqrt{2 \mu } \over  2 \pi \hbar } \bigg[ { 2 \sqrt{D - \tilde D}  \over  \tilde C}+ \int_{R_-^c }^{R_+^c  } { 1 \over \sqrt{ D - V(R)} } \d R \bigg] \nonumber
\end{eqnarray}

Formula (\ref{lRbernst_mod}) is a very simple one because all the three terms in formula~(\ref{gamma_form})
match the same $D-E$ behavior in 
formula (\ref{lRbernst_mod}). 

As $v_D-v$ does not depend on $R_+^ c$, expression~(\ref{lRbernst_mod}) shows that $\gamma$ should also be independent of $R_+^c$. As a consequence (see expression (\ref{gamma_form})), the non asymptotic part (from $R_-$ to $R_+^c$) should follow a  $(R_+^c)^{ {n+2
\over 2} } $ behavior. This is not fully satisfactory as the non asymptotic part should not depend on the value of $n$ which is a purely asymptotic parameter. This kind of trouble occurs whenever
 a cut-off is present in any theory. To prevent this caveat, we have hidden the cut-off $R_+^c$ inside the only parameter $\gamma$. Thus, the final formula (\ref{lRbernst_mod}) is no more depending on the cut-off value $R_+^c$.

 Considering the added term $\gamma (D-E)$ as a perturbation, it is possible to reverse the formula~(\ref{lRbernst_mod}) leading to our first improved formula:
\begin{equation}
D-E \approx \left(\frac{v_D-v}{ H_n^{-1} }\right)^{{2n \over n-2}}\left[1-{2n \over n-2}\gamma 
\left(\frac{v_D-v}{ H_n^{-1} }\right)^{{2n \over n-2}-1} \right]
\label{inv_LRB}
\end{equation}
The LeRoy-Bernstein formula is then
improved by simply adding one term depending on a single coefficient 
$\gamma$ that can be used as a parameter in a fit
procedure.
Will see in section~\ref{sec:test} how this formula improves 
the fit to the experimental energies values.

\section{Other multipole expansion coefficients. General formulas.}
\label{other_mult}

\subsection{General NDE formulas} 

To improve further the accuracy of the LeRoy-Bernstein formula,  we can add other multipole expansion coefficients as in formula~(\ref{Dv_mul}). The cut-off $R_+^c$ is then redefined so as to obtain with a typical $10\%$ accuracy:
\begin{equation}
V(R) \stackrel{ \scriptstyle R>R_+^c }{\approx} D+ {C_n \over R^n}+{C_m \over R^m}
\label{Dv_mul_cut_off}
\end{equation}

To be more general, let us  notice with LeRoy~\cite{LeRoy80b} that  the vibrational progression in $v_D-v$, the rotational constant $B_v$, the kinetic energy $T$, other BKW expressions of higher order, or higher rotational constants as $D_v$, 
can be derived from the integrals $I_{k,l}^{i,j}$:
\begin{equation}
I_{k,l}^{i,j} (E) = \int_{R_-(E)}^{R_+(E)}
 { \left( {\partial^i V(R) \over \partial R^i} \right)^j \over R^l(E-V(R))^{k+1/2} }  \d R
\label{int_tota}
\end{equation}
The derivative of $I_{k,l}^{i,j}$ relative to $E$  is equal to $-(k+1/2) I_{k+1,l}^{i,j}$.
 Thus, the NDE expression can be calculated only for the $k=0$ case: $I_{0,l}^{i,j}$. 
The method goes as follows.

As a first step, we separate the integral $\int_{R_-}^{R_+}$ in three parts  
$ \int_{R_-}^{R_-^c}+\int_{R_-^c}^{R_+^c}+\int_{R_+^c}^{R_+}$. 
The first integral is analytically calculated using the linear expression~(\ref{rep_lin}) for $V$. 
Using
$1/\sqrt{E-V(R)} \approx 1/\sqrt{D-V(R)}$ in the intermediate region, we compute the second integral as a simple number, independent of $E$. 
If we use the same assumption as in formula~(\ref{repu_branch}) the first two integrals, symbolically written as $I_l^{\na}(E)$, can be approximate by the number  $I_l^{\na}(D)$.
 In the asymptotic region (third integral), where $V$ is given by the polynomial multipole development~(\ref{Dv_mul_cut_off}), the numerator in the
wanted expression~(\ref{int_tota}) is just a $R$ polynomial expression, so it simplifies with the $R^l$ in the denominator. Finally, we need to calculate only a single expression: 
\begin{equation}
I_l^{\rm a} (E) = \int_{R_+^c}^{R_+(E)}
 { 1 \over R^l \sqrt{E-V(R)} }  \d R
\label{int_tot}
\end{equation}
 the subscript ``$^{\rm a}$'' is for ``asymptotic'' and we use similar notations as~\cite{LeRoy80}. We will only consider the case $0\leq l \leq 2$.

We simply have to follow the same kind of modification used in the previous section to calculate $I_{l}^{\rm a} (E)$.
The computation, detailed in appendix~\ref{app:gen_form}, is based on first order correction in
 $\alpha_c = {C_m / (R_+^c)^m \over C_n /(R_+^c)^n}
$. Consequently, using  the notations
$\beta={n+2-2l \over 2n}$ and  $\delta = \beta -\frac{m-n}{n}$, which can be null for the set of values $(n,m,l)=(4,5,2)$  or $(6,10,0)$ for instance, the complete (non-asymptotic) formula (\ref{final_formula_ann}) can be written in a more useful and compact form $I_l (E) = \int_{R_-(E)}^{R_+(E)}
 { 1 \over R^l \sqrt{E-V(R)} }  \d R $: 
\begin{eqnarray}
I_l &\approx& (-C_n)^{ -1/2 }\frac{ B(\beta,1/2)}{n} \left(  \frac{D-E}{-C_n} \right)^{-\beta}
+  \gamma_\delta + \label{final_formula_ter} 
\\
& &  \frac{(-C_n)^{ -1/2 }}{n} \frac{C_m}{C_n}  \left(  \frac{D-E}{-C_n} \right)^{-\delta} \left\{ \begin{array}{lr}
\beta B(\beta,\frac{1}{2})  & {\rm if \ } \delta<0 \\
\frac{1}{2} \ln  (D-E) & {\rm if \ } \delta=0 \\
( \delta - \frac{1}{2} )  B(\delta,\frac{1}{2}) & {\rm if \ } \delta>0 \\
\end{array} \right.
 \nonumber
\end{eqnarray}
where $  \gamma_\delta $ groups all the constants terms as the non-asymptotic ones and depends on $C_n, C_m, R_+^c$ ; for a precise value, usually not needed, see equation~(\ref{final_formula_ann}). $B(a,b) = \frac{\Gamma(a) \Gamma(b)}{\Gamma(a+b)}$
are
Euler  Beta functions.

Ignoring the $  \gamma_\delta $ 
term and choosing $C_m=0$, we recover the usual ("non"-improved) NDE formulas.

\subsection{Vibration}

We shall first develop an example of vibrational progression $v= {\sqrt{2 \mu} \over \pi \hbar} I_{-1,0}^{0,0} - 1/2$, it is nothing else than formula (\ref{cond_quant_BKW}). 
The NDE development for $v_D-v$ is  given by integrating the differential equation:
$
{\d v \over \d E} = {\sqrt{2 \mu} \over 2 \pi \hbar} I_0
$, 
leading to:
\begin{eqnarray}
 v_D-v  &\approx& {\sqrt{2 \mu} \over 2 \pi \hbar} \Big[ (-C_n)^{ \beta -1/2 }  \frac{B(\beta,1/2)}{n} \frac{(D-E)^{1-\beta}}{1-\beta} \nonumber \\
  & & + \tilde \gamma_\delta (D-E)  \label{vd_final} \label{fit_LRB}\\
& &  + \frac{(-C_n)^{ \delta -1/2 }}{n} \frac{C_m}{C_n}   \frac{(D-E)^{1-\delta} }{1-\delta}  \left\{ \begin{array}{lr}
\beta B(\beta,\frac{1}{2})  & {\rm if \ } \delta<0 \\
\frac{1}{2} \ln  (D-E) & {\rm if \ } \delta=0 \\
( \delta - \frac{1}{2} )  B(\delta,\frac{1}{2}) & {\rm if \ } \delta>0 \\ \end{array} \right. \Big] \nonumber
\end{eqnarray}
where, for instance,  $
\tilde \gamma_\delta = {\sqrt{2 \mu} \over 2 \pi \hbar}  \gamma_\delta$ for $\delta \not = 0$.

\subsection{Rotation}
Using the expression of  the averaged
semi-classical probability of presence (e.g. see~\cite{LandauMQ}) for the radial wavefunction $\psi_v$:$
  \omega \mu \over \pi 
 \sqrt{ 2 \mu (E - V(R))}
  $, and expression for
 $\omega$ using formula~(\ref{LRBERN}), 
 we 
 derive the NDE analytical expression
for the rotational constant 
$B_v = {\hbar^2 \over 2 \mu} \langle \psi_v | {1 \over R^2} | \psi_v \rangle$~\cite{LeRoy72}:
\begin{equation}
B_v = {\hbar^2 \over 2 \mu} {I_2 \over I_0}
\end{equation}
 It is well known that this formula is less accurate than the formula~(\ref{lRbernst}) for the vibrational progression. Indeed,
we had to neglect in the non-asymptotic region (i.e. for small $R$ values) the
$1/R^2$ term in the $B_v$ calculation.
Our improved formula (\ref{final_formula_ter}) should also help to solve  this point.

\subsection{Kinetic energy}

 Similarly, we can compute the average kinetic energy
$\langle T \rangle = \langle \psi_v | E-V(R) |
\psi_v \rangle $~\cite{Stwalley73}:
\begin{equation}
 \langle T \rangle  =  {1 \over 2} { (v +1/2) \over 
{\d v \over \d E} (E) }
= { \pi \hbar \over \sqrt{2 \mu} } {v +1/2 \over I_0}
= {I_{-1,0}^{0,0} \over I_0}
 \label{kin_equ}
\end{equation}
This problem is indeed crucial
 in photoassociation. During its
de-excitation, the kinetic energy of the PA excited molecule is transferred to
the two free atoms. They can then leave the magneto-optical trap, if their
speed is sufficient, leading to a detectable signal.

\section{Experimental comparison}
\label{sec:test}

The validity of the usual LeRoy-Bernstein formula~(\ref{inv_LRB}) with $\gamma=0$
 has been studied~\cite{LeRoy80b,LeRoy80} but not extensively.  We are going to study the own performances
 of  our formula.

\subsection{Multipolar development for dialkalis molecules}

We shall not present here a complete review of the former applications of the LeRoy-Berstein law to photoassociation spectroscopy. Although we won't give an overview of the theoretical study of the Hund  case (c) long-range potential curve, we want to give a brief introduction to the subject in order to be able to compare the calculated coefficients $C_n$ and $C_m$ with the modified NDE formulas.
A detail introduction to the Hund  case (c) potential curve calculation is given in appendix~\ref{long_range}. Our final goal is to obtain the $C_n$ and $C_m$ leadings coefficients for the long-range states and to be able to take into account all terms needed to reach $1\%$ accuracy in the $C_n$ value. We will focus on this section
 on alkali homonuclear molecules dissociating toward $ns+n'p$ asymptote.

Hund  case (a) potential curve dissociating toward the $ns+n'p$ asymptotic limit
of the two identical alkali atoms leads to the following multipolar expansion:
\begin{eqnarray}
V_{ ^3 \Sigma_u^+ }  = V_{ ^1 \Sigma_g ^+ } & = &  2 {C_3\over R^3} + {C_6^{\Sigma} \over R^6} + { C_8^{\Sigma_s} \over R^8} + \ldots \label{dev_cas_a} \\
 V_{ ^3 \Pi_g }  = V_{ ^1 \Pi_u  } & = &   {C_3 \over R^3} + {C_6^{\Pi}\over R^6} + {C_8^{\Pi_a} \over R^8} + \ldots \nonumber \\
V_{ ^3 \Pi_u }  = V_{ ^1 \Pi_g  } & = &  - {C_3 \over R^3} + {C_6^{\Pi}\over R^6} + {C_8^{\Pi_s} \over R^8} + \ldots  \nonumber \\
V_{ ^3 \Sigma_g^+ }  = V_{ ^1 \Sigma_u ^+ } & = &  - 2 {C_3 \over R^3} + {C_6^{\Sigma}\over R^6} + {C_8^{\Sigma_a} \over R^8} + \ldots  \nonumber
\end{eqnarray}
where coefficients are given in the table~\ref{coef_theo} for the cesium case.

 \begin{table}
$$   \begin{array}{|l|l|}\hline
C_3 & 9.997 (23) \\ \hline
C_6^\Sigma & -17390 \\\hline
C_6^\Pi & -11830 \\\hline
C_8^{\Sigma_a} & -5040000 \\\hline
C_8^{\Sigma_s} & -16560000 \\\hline
C_8^{\Pi_a} & -2256000 \\\hline
C_8^{\Pi_s} & -913100 \\\hline
   \end{array}
$$
 \caption{Multipolar coefficient in formula~ (\ref{dev_cas_a}) for the cesium case at the $6s+6p$ asymptote.
 $C_3$ is closely related to the dipole matrix element~(\ref{new_c_3_def}) and therefore the atomic lifetime (formula~(\ref{ato_lifetime})). The $C_3$ value given here is extracted from atomic lifetime measurement~\cite{Rafac99}.
$C_6$ and $C_8$ coefficients were theoretically calculated~\cite{Marinescu95}.}
\label{coef_theo}  
 \end{table}

\begin{figure}[ht]
\begin{center}
\resizebox{\textwidth}{0.5\textwidth}{\includegraphics{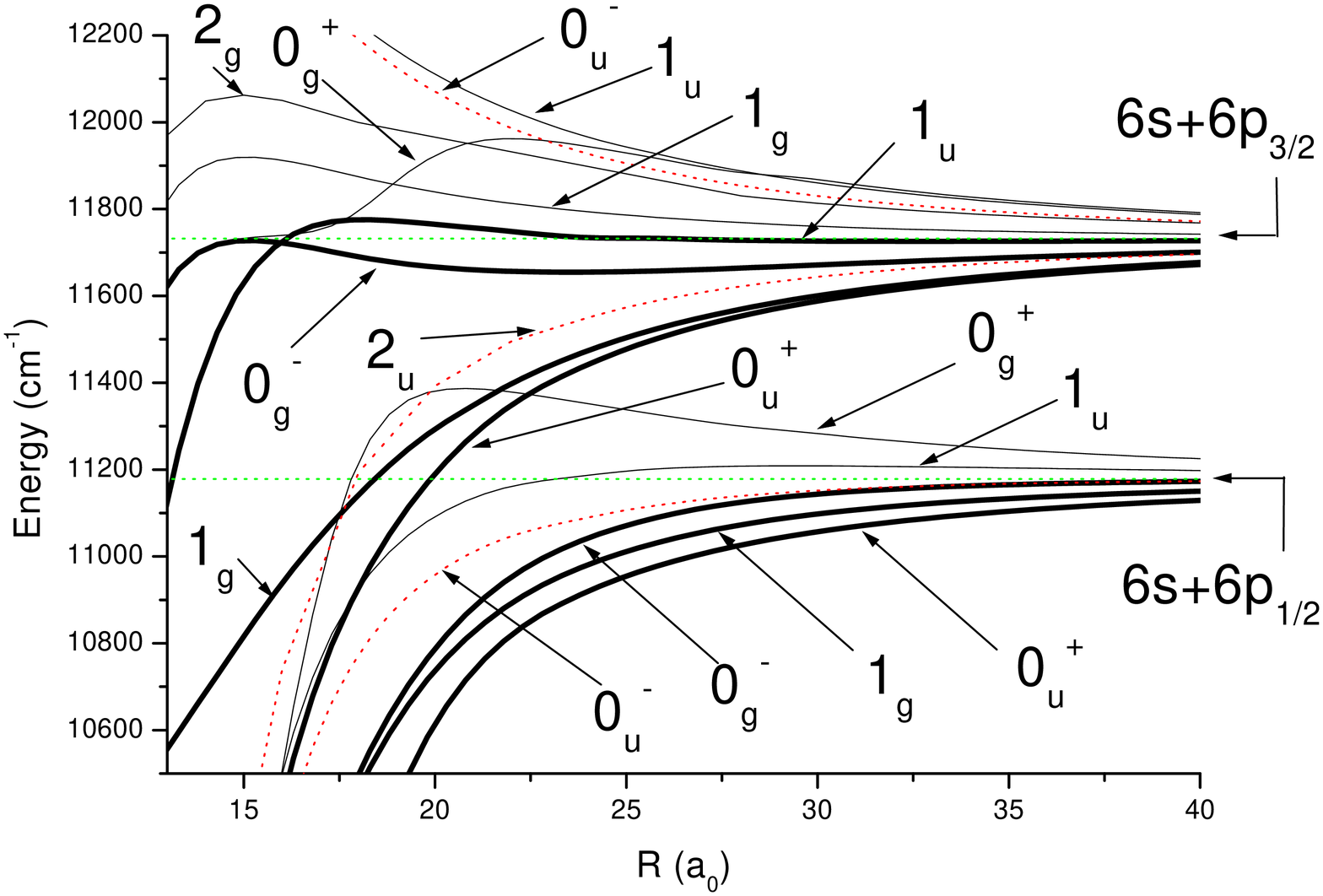}}
\end{center}
\caption{Hund case (c) potential curves for the $6s+6p_j$ dissociation limits in the cesium case. The $7$ attractive allowed curves (broad solid lines) for a dipolar transition, as the photoassociation transition from $6s+6s$ atoms, are shown with repulsive (tiny solid lines) and forbidden  ones (dotted lines). The curves are issue of a matching between short-range {\em ab-initio} calculation~\cite{spiess89} and long range ones given in the formula~(\ref{dev_cas_a}) and the table~\ref{coef_theo}.}
\label{pot_curve}
\label{fig:hundc}
\end{figure}

The Hund  case (c) potential curve (see figure~\ref{pot_curve}) are obtained after diagonalization of the matrix
$M+\delta M$. $M$ is given in
table~\ref{table_etat(c)}. $\delta M$ is a correction matrix
 given in table~\ref{corr_matrix} which has to my knowledge never been published before. 
After the diagonalization of such matrices we calculate the multipolar expansion, i.e. the power series by respect to $1/R$. Results are summarized in table~\ref{asymp_coef}. Let us mention that
 the "real"  accurate expansion will contain other terms coming from retardation ($1/R$ dependence), Coriolis ($1/R^2$ dependence) or spin-spin ($1/R^3$ dependence) effects. But these effects  are negligible compared to  the $C_n/R^n$ term in the multipolar expansion for the range of internuclear distance ($20-200\,a_0$)  we are working with. Nevertheless their contributions are evaluated in appendix \ref{long_range} with the new relativistic lifetime correction $\delta M$. 

\begin{table}
\begin{eqnarray*}
V_{\pm 2_u} & = & \frac{A}{2} - {C_3 \over R^3} + {C_6^{\Pi_s}\over R^6}   \\
V_{\pm 2_g} & = &  \frac{A}{2} + {C_3 \over R^3} + {C_6^{\Pi_a}\over R^6}  \\
V_{\pm 1_u} & = &  \left\{ 
\begin{array}{l}
\frac{{A}}{2} + 
   \frac{\left( 2 + {\sqrt{7}} \right) \,{C_3}}{3\,R^3} + 
   \frac{9\,{A}\,\left( 2\,\left( -1 + {\sqrt{7}} \right) \,
          {C_6^\Pi} + \left( 2 + {\sqrt{7}} \right) \,
          {C_6^\Sigma} \right)  + 
      2\,\left( -10 + 7\,{\sqrt{7}} \right) \,{{C_3}}^2\,
       {\left( 1 + \epsilon  \right) }^2}{27\,{\sqrt{7}}\,
      {A}\,R^6} 
\cr
-{A} + \frac{2\,{C_3}\,
      {\left( 1 + \epsilon  \right) }^2}{3\,R^3} + 
   \frac{9\,{A}\,\left( 2\,{C_6^\Pi} + 
         {C_6^\Sigma} \right)  - 
      28\,{{C_3}}^2\,{\left( 1 + \epsilon  \right) }^2}{27\,
      {A}\,R^6} 
\cr
 \frac{{A}}{2} - \frac{\left( -2 + {\sqrt{7}} \right) \,
      {C_3}}{3\,R^3} + 
   \frac{9\,{A}\,\left( 2\,
          \left( {C_6^\Pi} - {C_6^\Sigma} \right)  + 
         {\sqrt{7}}\,\left( 2\,{C_6^\Pi} + {C_6^\Sigma} \right)  \
\right)  + 2\,\left( 10 + 7\,{\sqrt{7}} \right) \,{{C_3}}^2\,
       {\left( 1 + \epsilon  \right) }^2}{27\,{\sqrt{7}}\,{A}\,
      R^6} 
\cr
\end{array}
\right.
\\
V_{\pm 1_g} & = &  \left\{ 
\begin{array}{l}
\frac{{A}}{2} + 
   \frac{\left( -2 + {\sqrt{7}} \right) \,{C_3}}{3\,R^3} + 
   \frac{9\,{A}\,\left( 2\,
          \left( {C_6^\Pi} - {C_6^\Sigma} \right)  + 
         {\sqrt{7}}\,\left( 2\,{C_6^\Pi} + {C_6^\Sigma} \right) 
 \right)  + 2\,\left( 10 + 7\,{\sqrt{7}} \right) \,{{C_3}}^2\,
       {\left( 1 + \epsilon  \right) }^2}{27\,{\sqrt{7}}\,
      {A}\,R^6} 
\cr
 -{A} - \frac{2\,{C_3}\,
      {\left( 1 + \epsilon  \right) }^2}{3\,R^3} + 
   \frac{9\,{A}\,\left( 2\,{C_6^\Pi} + 
         {C_6^\Sigma} \right)  - 
      28\,{{C_3}}^2\,{\left( 1 + \epsilon  \right) }^2}{27\,
      {A}\,R^6} 
\cr
\frac{{A}}{2} - \frac{\left( 2 + {\sqrt{7}} \right) \,
      {C_3}}{3\,R^3} + 
   \frac{9\,{A}\,\left( 2\,\left( -1 + {\sqrt{7}} \right) \,
          {C_6^\Pi} + \left( 2 + {\sqrt{7}} \right) \,
          {C_6^\Sigma} \right)  + 
      2\,\left( -10 + 7\,{\sqrt{7}} \right) \,{{C_3}}^2\,
       {\left( 1 + \epsilon  \right) }^2}{27\,{\sqrt{7}}\,{A}\,
      R^6} 
 \cr
\end{array}
\right.
\\
V_{0_u^+} & = &\left\{ 
\begin{array}{l}
-A - \frac{4\,{C_3}\,
      {\left( 1 + \epsilon  \right) }^2}{3\,R^3} + 
   \frac{9\,A\,\left( 2\,{C_6^\Pi} + 
         {C_6^\Sigma} \right)  - 
      4\,{{C_3}}^2\,{\left( 1 + \epsilon  \right) }^2}{27\,
      A\,R^6} 
\cr
  \frac{A}{2} - \frac{5\,{C_3}}{3\,R^3} + 
   \frac{9\,A\,\left( {C_6^\Pi} + 
         2\,{C_6^\Sigma} \right)  + 
      4\,{{C_3}}^2\,{\left( 1 + \epsilon  \right) }^2}{27\,
      A\,R^6} 
\cr
\end{array}
\right. \\
V_{0_g^+} & = & \left\{ \begin{array}{l}
-A + \frac{4\,{C_3}\,
      {\left( 1 + \epsilon  \right) }^2}{3\,R^3} + 
   \frac{9\,A\,\left( 2\,{C_6^\Pi} + 
         {C_6^\Sigma} \right)  - 
      4\,{{C_3}}^2\,{\left( 1 + \epsilon  \right) }^2}{27\,
      A\,R^6} 
\cr
\frac{A}{2} + \frac{5\,{C_3}}{3\,R^3} + 
   \frac{9\,A\,\left( {C_6^\Pi} + 
         2\,{C_6^\Sigma} \right)  + 
      4\,{{C_3}}^2\,{\left( 1 + \epsilon  \right) }^2}{27\,
      A\,R^6} 
\cr
\end{array}
\right. \\
V_{0_u^-} & = & \left\{ 
\begin{array}{l}
-A + \frac{A\,
       \left( 2\,{C_6^\Pi} + {C_6^\Sigma} \right)  - 
      4\,{{C_3}}^2\,{\left( 1 + \epsilon  \right) }^2}{3\,
      A\,R^6} + \frac{2\,{C_8^{\Pi_s}} + {C_8^{\Sigma_s}}}
    {3\,R^8} 
\cr
 \frac{A}{2} + \frac{{C_3}}{R^3} + 
   \frac{A\,\left( {C_6^\Pi} + 2\,{C_6^\Sigma} \
\right)  + 4\,{{C_3}}^2\,{\left( 1 + \epsilon  \right) }^2}{3\,
      A\,R^6} + \frac{{C_8^{\Pi_s}} + 2\,{C_8^{\Sigma_s}}}
    {3\,R^8} 
\cr
\end{array}
\right. \\
V_{0_g^-} & = & \left\{ 
\begin{array}{lc}
-A + \frac{A\,
       \left( 2\,{C_6^\Pi} + {C_6^\Sigma} \right)  - 
      4\,{{C_3}}^2\,{\left( 1 + \epsilon  \right) }^2}{3\,
      A\,R^6} + \frac{2\,{C_8^{\Pi_a}} + {C_8^{\Sigma_a}}}
    {3\,R^8} 
\cr
\frac{A}{2} - \frac{{C_3}}{R^3} + 
   \frac{A\,\left( {C_6^\Pi} + 2\,{C_6^\Sigma} \
\right)  + 4\,{{C_3}}^2\,{\left( 1 + \epsilon  \right) }^2}{3\,
      A\,R^6} + \frac{{C_8^{\Pi_a}} + 2\,{C_8^{\Sigma_a}}}
    {3\,R^8} 
\cr
\end{array}
\right. \\
\end{eqnarray*}
\caption{Two main term in the multipolar expansion in $1/R$ for Hund  case (c) potential curve.
The dissociation is taken at $ns+n'p$ limit. Then, levels starting with $A/2$ dissociate toward $ns + n'p_{3/2}$, while the $ -A$ levels dissociate toward $ns+n'p_{1/2}$.}
\label{asymp_coef}
\end{table}

\subsection{Testing the improved LeRoy-Bernstein formula with the Cs$_2$ $0_g^- (6s+6p_{3/2})$ state.}
\label{cas_og}

Our group(~\cite{Fioretti99,Comparat00,Dion}) obtained experimental photoassociation spectra with an accuracy of $150\,$MHz for all the seven
 allowed states (see figure~\ref{pot_curve}).
We will focus here on the spectrum, from $v=0$ ($D-E_v \approx 77\,$cm$^{-1}$) to $v=133$ ($D-E_v \approx 0.4\,$cm$^{-1}$), of the external well of the 
 $0_g^- (p_{3/2})$ state.  To test our formulas we will use the data extracted from a RKR study and published in \cite{cm:fioretti99}.
Using table~\ref{coef_theo} and \ref{asymp_coef}, we shall take 
 $n=3,m=6, l=0$ and $ C_n\approx -10\, e^2 a_0, C_m \approx 65000\,e^2 a_0^5$. 
This is one 
of the most ruthless case to test a LeRoy-Bernstein  type formula because almost
 all the assumptions used in its derivation are wrong or could be discussed. The improved version
 shall then be needed.

 Firstly, $\delta<0$ (and $C_m>0$) lead to problems that LeRoy in reference \cite{LeRoy80} had. This is due to the fact that $T_{0,1}^{3,6}(0)=\infty $ in formula (\ref{tl1}). Our derivation avoid them using $T_{0,1}^{3,6}(y^n)$ which is finite.

Secondly $n$ is small and we use a "pure long range state" where $R$ is always large. Therefore, $x^n = (R/R_+)^n$ is not so small and
 keeping only
 the first order, as done in the "usual" LeRoy-Bernstein derivation, in the series~(\ref{first_improv}) might be not accurate enough.

Thirdly, the term $C_m \approx 65000\,e^2 a_0^5$ is larger than in usual Hund case
 (a) potential curve, therefore the $C_m/R^m$ correction term is also large. Furthermore $R_+^c$ is of greater value and  $y=R_+^c/R_+$ is not small (see for instance formula~(\ref{form_inte})). 

Fourthly, the inner wall is smooth and less steep than usual (see figure \ref{fig:hundc}). Consequently a small phase is accumulated on the inner wall (see formula (\ref{repu_branch})).

In the fifth place, the potential curve is only $80\,$cm$^{-1}$ deep. So,
$ {1 / \sqrt{E - V(R)} }$ is quite large and the non-asymptotic part, defined in 
expression~(\ref{asy_non_asym}), is therefore important. For the same reason, the assumption: $ 1 / \sqrt{E - V(R)} \approx  1 / \sqrt{D - V(R)} $ in the intermediate region can also be wrong.

For $R_+\approx 40\,a_0$, {\it i.e.} $D-E \approx 30 \,$cm$^{-1}$, 
 the next asymptotic coefficient $C_8/R^8$ is already $4\%$ of $C_6/R^6$ (it reaches $10\%$ for $R\approx 25\,a_0$).
Therefore a choice of $R_+^c \approx 35-40\, a_0$ is probably good enough to obtain
$|\alpha_c| <0.1$ and a non-asymptotic part not too large.
As mentioned in section~\ref{pb_asym},
 we also need
$R_+/R_+^c$ large enough
to get a good precision in our fit. Concerning the percent accuracy goal, the discussion following formula~(\ref{trois_equa}) has indicated a  restriction
for the fit
 at $R_+> 1.6 R_+^c \approx 55\,a_0$ corresponding to $D-E < 12\,$cm$^ {-1}$. 

For a physical insight,  
 we  give in table~\ref{ordre_grandeur} some typical values  for all the terms present in formula~(\ref{vd_final}). 
The potential curved used to numerically evaluate all the terms in table~\ref{ordre_grandeur} including the term
\begin{equation}
\tilde \gamma_\delta = {\sqrt{2 \mu} \over 2 \pi \hbar}  
\left( I_l^{\na}(D) -  (-C_n)^{ -1/2 } \frac{(R_+ ^c)^{n \beta } }{n\beta} + \frac{1}{2} (-C_n)^{ -1/2 } \frac{C_m}{C_n} \frac{(R_+ ^c)^{n \delta } }{n\delta}  \right)
\label{term_LRB}
\end{equation}
is simply the 
diagonalization of the $0_g^-$ matrix describe above where the $C_8$ and $\epsilon$ coefficients are assumed to be zero.
These results confirm the well known fact the "usual" LeRoy-Bernstein formula  (with the sole term $(D-E)^{1-\beta}$) won't gives results better than one percent. This also confirms that the "usual" LeRoy-Bernstein formula is by chance much more accurate than it should be because the non-asymptotical parts  $I_l^\na (D)$ in $\tilde \gamma_\delta$  almost perfectly canceled with the other terms such as $ \hat \gamma_\beta$.

\begin{table}
$$
\begin{array}{|c|c|c|c|}
\hline
D-E ({\rm cm}^{-1}) & 30 & 10 & 1 \\
\hline
\hline
 (D-E)^{1-\beta} & 172 &143 &97  \\
 \hline
 \hat \gamma^\na (D)={\sqrt{2 \mu} \over 2 \pi \hbar} I_l^{\na}(D)  & 12.7&4.2 & 0.4 \\
\hline
 \hat \gamma^\na (E)={\sqrt{2 \mu} \over 2 \pi \hbar} I_l^{\na}(E)  & 16.2&4.5 & 0.4 \\
\hline
 \hat \gamma_\beta & -9.8& -3.3& -0.3 \\
 \hline
 \hat \gamma_\delta & 3.7& 1.2& 0.1 \\
\hline
  (D-E)^{1-\delta} & -1.8& -0.5&  -0.03\\
\hline
 O(y^n) & -0.7& -0.1&  -0.001 \\
\hline
\end{array} $$
\caption{Estimation, for several $D-E$ values (in cm$ ^{-1}$), of the terms involved in $v_D-v$ formula~(\ref{fit_LRB}). We use the analytical $0_g^-(p_{3/2})$ potential curve with $C_8=\epsilon=0$, where 
we have chosen $R_+^c = 35\,a_0$.
Terms are labeled by the $D-E$ power in the formula~(\ref{fit_LRB}). The three separate terms in formula (\ref{term_LRB}) are respectively noted $\hat \gamma^\na, \hat \gamma_\beta, \hat \gamma_\delta$. We also evaluate the contribution from the neglected $O(y^n)$ term in equation (\ref{tl0}) which is the second term in formula (\ref{first_improv}).}
\label{ordre_grandeur}
\end{table}

In the non-asymptotic part  we have made the approximation, as in formula~(\ref{non_asy_eq}),
$
I_l^\na (E) \approx I_l^\na (D) 
$ which, for instance with $D-E \approx 10\,$cm$^{-1}$, leads to an accuracy of the $v_D-v$ value of about $0.3$. This is of similar importance than other  contributions listed in table~\ref{ordre_grandeur}. 
As a consequence, it is useless take into account contributions smaller than $0.3$. 
Therefore, this means  that in order to improve our formula, we shall not incorporate second order terms (such as $O(y^n)$)
but  we shall rather have to  take into account more carefully the non-asymptotic part. This can not be done without adding other unknown parameters in the development as ${\d \tilde \gamma_\delta \over \d E}$ or $R_+^c$-dependent terms and without keeping $E$ in formula~(\ref{repu_branch}). In a sense our theory is the best with only one single unknown parameter added $\tilde \gamma_\delta$ to $v_D,D,C_n$ and $C_m$.  

Our theory, see expression (\ref{vd_final}) yields $v_D-v = f(D-E)$, but we would prefer to adjust the theory to the experimental energies, i.e. to fit using
$D-E=f^{-1} (v_D-v)$. 
Table~\ref{ordre_grandeur} indicates the formula~(\ref{fit_LRB})
is largely dominated by the first term $ H_n^{-1}  (D-E )^{1-\beta}$ and that we could safely, except for very large $D-E$ values, use the only first order inversion procedure used for instance to
derive formula~(\ref{inv_LRB}) to find an accurate enough  general reversed formula:

\begin{eqnarray}
D-E  &\approx&  \left(  \frac{v_D-v}{ H_n^{-1} } \right)^{\frac{1}{1-\beta}}
\big[1- \frac{1}{1-\beta} \frac{1 }{v_D-v} \Big(
\tilde \gamma_\delta \left(  \frac{v_D-v}{ H_n^{-1} } \right)^{\frac{1}{1-\beta}} +
\label{LRB_final}
\\
& & 
\frac{(-C_n)^{ \delta -1/2 }}{n} \frac{C_m}{C_n}   \frac{1}{1-\delta} 
\left(  \frac{v_D-v}{ H_n^{-1} } \right)^{\frac{1-\delta}{1-\beta}}
 \left\{ \begin{array}{lr}
\beta B(\beta,\frac{1}{2})  & {\rm if \ } \delta<0 \\
\frac{1}{2} \ln  \left(  \frac{v_D-v}{ H_n^{-1} } \right)^{\frac{1}{1-\beta}}
 & {\rm if \ } \delta=0 \\
( \delta - \frac{1}{2} )  B(\delta,\frac{1}{2}) & {\rm if \ } \delta>0 \\
\end{array} \right.
\Big) \Big] \nonumber
\end{eqnarray}

This formula gives an explanation for the origin of the Pade  coefficients used in "usual" NDE formula \cite{cm:reza79}, as long as an explanation for their values.
Indeed, Pade formulas assume a mathematical, without {\em a-priori} a physical meaning, polynomial quotient expansion in $v_D-v$ for physical value as vibrational series or rotational constants \cite{LeRoy94}.
Our formula leads directly to such polynomial expansion and gives physical interpretation for the coefficients.

In table~\ref{fit_resu}, we present fit results  (done with Mathematica software, with 100 iterations in the non linear fitting procedure) for all the vibrational levels and for the vibrational $-30,-10,-5$ and$-2\,$cm$^{-1}$ of the $0_g^-$ state. It should be remembered that we use data from RKR computation  where $C_3(0_g^-)$ was kept  fixed at $-10.47 e^2 a_0^2$~\cite{cm:fioretti99}. The accuracy is also much better than the experimental one because we use a NDE theory to fit the NDE-RKR calculated levels and not the experimental levels.
The table~\ref{fit_resu} shows an improvement when using our formula, as opposed to the usual LeRoy-Bernstein one.
Our method is much more stable than the usual ones when the fitted energy range changes. Thus we are able to extract
$C_3\approx -10.5,v_D\approx 214.8, D\approx -0.09\,$cm$^{-1}$ which are very close to values found by a complete
NDE-RKR analysis~\cite{Fioretti99} $C_3=-10.47, v_D\approx 214.6, D\approx -0.09\,$cm$^{-1}$. 

Our method seems suitable for extracting asymptotic coefficients at the percent accuracy level. The method could be accurate enough to need the corrections factors as the $\delta M$ matrix as the $\epsilon$ one which has never been used up to know.
The method can also be applied to other states as the $0_g^- (p_{1/2})$ or the $1_g (p_3/2)$ ones. The $0_g^- (p_{1/2})$ state has a $1/R^6$ asymptotic behavior (see table \ref{asymp_coef}) and then should gives information concerning the next multipolar coefficients $C_6^{\Sigma}$ and $C_6^{\Pi}$. The $1_g (p_3/2)$ leads to $n=3$ and $m=8$ in the cesium case because, see table \ref{asymp_coef} and value given in table \ref{coef_theo}, the $1/R^6$ term is accidentally very small and therefore negligible comparing to the $1/R^8$ one.

\begin{table}
\begingroup
$$
\begin{array}{|l||c|c|c|c|c|c||c|}
\hline 
 \hbox{method}  & {\rm range} & D & 
C_3(0_g^-) & v_D &  \tilde \gamma_\delta & C_6 & \sigma^{\rm fit}\\
\hline
(\ref{leroybern}) & {\rm all} & 0.85&-17.94&245.1&0&0&9879 \\
\hline
&-30<E_v& 0.16 & -14.10 & 228.7 &0&0&1906  \\
\hline
& -10& -0.03&-12.16&220.6&0&0&230 \\
\hline
& -5& -0.06&-11.60&218.4&0&0&59 \\
\hline
& -2& -0.08&-11.12&216.6&0&0&11 \\
\hline
\hline
(\ref{inv_LRB}) & {\rm all} & 0.06&-12.42&222.6&14686&0&1272 \\
\hline
&-30& -0.06& -11.27 & 221.7 &20142&0&151  \\
\hline
& -10& -0.084&-10.79&215.7&26378&0&14 \\
\hline
& -5&  -0.088&-10.68&215.3&29544&0&9 \\
\hline
& -2& -0.091&-10.51&214.8&37070&0&9 \\
\hline
\hline
(\ref{LRB_final}) & {\rm all} & -0.080&-10.71&215.6&56034&-1.3 10^7&61 \\
\hline
&-30& -0.089 & -10.55 & 215.0 &62565&-1.5 10^7&10  \\
\hline
& -10& -0.090&-10.49&214.8&69581&-1.7 10^7&9 \\
\hline
& -5& -0.090&-10.48&214.7&73770&-1.9 10^7&9 \\
\hline
& -2& -0.091&-10.52&214.8&37019&-25061&10 \\
\hline
\hline
\end{array}
$$
\endgroup
 \caption{Value of the coefficients $D,C_3(0_g^-),v_D, \tilde \gamma_\delta ,C_6 $ for the $0_g^- (6s+6p_{3/2})$ cesium state. the numerical fit was done using formula~(\ref{LRB_final}) with $k=5$ parameters for different set of $N$ vibrational states determined by the range of  available $E_v$ values. When $\tilde \gamma_\delta=C_6=0$ ($k=3$), we recover  the usual LeRoy-Berstein formula~(\ref{leroybern}). When $C_6=0$ ($k=4$),  we recover  the first improved formula~(\ref{inv_LRB}). $\sigma^{\rm fit}=\frac{1}{N-k}\sqrt{
 \sum_{j=1}^N (E_v - E_v^{\rm fit})^2
 }$  is given in MHz units so as to be compared to the experimental accuracy $\sigma=150\,$MHz.}
\label{fit_resu}
 \end{table}

\section{Conclusion}
We have derived general improved NDE expansion formulas (\ref{final_formula_ter}), 
including LeRoy-Bernstein  one (\ref{LRB_final}), leading to a better accuracy in the determination of the asymptotic coefficients.

Such expressions can be useful for further
photoassociation experiments to extract the asymptotic coefficient $C_3$ or $C_6$ and the atomic lifetime.

The method gives also a physical meaning for the Pade coefficients used in usual NDE formula. The method could then be used as a starting guide for Pade coefficients calculation.
Furthermore our theory includes, as physical parameters, the two leading
multipolar coefficients $C_n$ and $C_m$. 
 The single added parameter $\tilde \gamma_\delta$ contains information on the repulsive branch, the intermediate internuclear distance
behavior and analytical calculation of the non-asymptotic part of the vibrational phase in a $C_n/R^n$ potential curve.
 We have shown it is not reasonable, without adding another parameters, to develop further than we did the approximation in series development of the analytically known asymptotical part. 

The author thank  C. Amiot, H. Blanchard, O. Dulieu, D. Hardin and N. Vanhaecke 
for many helpful discussions.

$^{\dag }$Laboratoire Aim{\'{e}} Cotton is associated with Universit{\'{e}}
Paris-Sud. web site: $www.lac.u$-$psud.fr$

\appendix

\section{Derivation of the improved LeRoy-Bernstein formula}
\label{app:gen_form}
We will detail derivation starting from formula~(\ref{int_tot}).

Using similar notations as~\cite{LeRoy80}, we define:
\begin{eqnarray}
x &=& R/R_+ \nonumber \\
y & = &  {R_+^c \over R_+ }  \nonumber \\
\alpha (R) &= &{C_m / R^m \over C_n /R^n} = \alpha_c (y/x)^{m-n} \ {\rm where }\\
\alpha_c  &= & \alpha (R_+^c) \ {\rm and} \label{def_alpha} \\
\alpha & = & \alpha (R_+) = \alpha_c y^{m-n} \label{alpha_y} 
\end{eqnarray}
 $C_m/R^m \ll C_n /R^n$ in the asymptotic region ($R>R_+^c$) is equivalent as saying that $\alpha_c \ll 1$. This inequality can be assured by a right choice of $R_+^c$. We shall then   use a development in Taylor  series about $\alpha_c=0$. We  define the "zero"-order parameters:
\begin{eqnarray}
y_0^n & =& \frac{D-E}{(-C_n)/(R_+^c)^n} \label{def_y0} \\
\alpha_0 & = & \alpha_c  y_0^{m-n} = \frac{C_m}{C_n} \left( \frac{D-E}{-C_n} \right)^{\frac{m-n}{n}} \label{def_alpha0}
\end{eqnarray}
 Then expression (\ref{Dv_mul_cut_off}) becomes: 
$$
V(R)  \stackrel{ \scriptstyle R>R_+^c }{\approx}  D+ {C_n \over R^n} ( 1 + \alpha(R) ) 
$$
and equation $E=V(R_+)$ leads to the following implicit equation for $y$:
\begin{equation}
y^n(1+\alpha_c y^{m-n})  =  y_0^n \label{eqn_y0}
\end{equation}
Using $\beta={n+2-2l \over 2n}>0$,
equation~(\ref{int_tot}) is easily written as:
\begin{equation}
 I_l^{\rm a}  =  (-C_n)^{ -1/2 }   \left( \frac{R_+^c}{y} \right)^{n \beta} 
  \int_y^1 \frac{  x^{n/2-l} }{\sqrt{1 -x^n}}\frac{1}{\sqrt{1 + \alpha(R) \frac{ 1- x^m}{1-x^n}}} 
 \d x
\label{equ_fin}
\end{equation}
The derivation is quite similar to that of equation (\ref{asy_non_asym}).
As $0 \leq x \leq 1$,  we have $1 \leq  \frac{ 1- x^m}{1-x^n} \leq \frac{m}{n}$. As a consequence, the second term is a perturbative one (because $\alpha(R) \leq \alpha_c \ll 1$), so we can expand the inverse of the square root in its  converging Taylor series: 
\begin{equation}
 I_l^{\rm a}  =  (-C_n)^{ -1/2 }   \left( \frac{(R_+^c)^n}{y^n} \right)^{\beta} 
\sum_{k=0}^\infty  \big( \begin{array}{c} -1/2 \\ k \end{array} \big)
\left( \alpha_c y^{m-n} \right)^k
T_{l,k}^{n,m}(y^n)
\label{equ_fin_u}
\end{equation}
 where $ \big( \begin{array}{c} -1/2 \\ k \end{array} \big) = \frac{(-1/2)(-1/2-1) \ldots (-1/2-k+1)}{k !}$, using $u=x^n$ and:
\begin{equation}
T_{l,k}^{n,m}(y^n) = \frac{1}{n}  \int_{y^n}^1 \frac{  u^{\beta - 1 - k \frac{m-n}{n} } }{ \sqrt{1 -u}} 
\left( \frac{1-u^{m/n}}{1-u} \right)^k \d u
\label{tl1_def}
\end{equation}
We use  a similar notation as in LeRoy's paper~\cite{LeRoy80}. However, contrary to his computation,
where  $y=0$ is fixed before
expanding into  Taylor series in $\alpha$ (not $\alpha_c$ as we did), thus leading to diverging integrals $T_{l,k}^{n,m}(0)$, we 
isolate here the dependence in a constant term $\alpha_c$ and
take into account the fact that $\alpha$ is $y$ dependent. Indeed formula (\ref{alpha_y}) indicates $\alpha \rightarrow 0$ when $y \rightarrow 0$. Furthermore, LeRoy can not deal with $\alpha<0$ (i.e. $C_m<0$), as we do in section \ref{cas_og}.

Equation~(\ref{form_inte}) 
has shown that 
\begin{equation}
T_{l,0}^{n,m}(y^n) = \frac{B(\beta,1/2)}{n} - \frac{ y^{n \beta}}{ n\beta} (1 + O (y^n)),
\label{tl0}
\end{equation}
with the Euler Beta functions:
\begin{eqnarray}
B(a,b) &=& \int_0^1 u^{a-1} (1-u)^{b-1} \d u  \ ; \ a,b>0 \nonumber \\
& = & \frac{\Gamma(a) \Gamma(b)}{\Gamma(a+b)} = \frac{a+b}{a} B(a+1,b)
\label{beta_euler}
\end{eqnarray} 
Similarly we have to expand  the integral (\ref{tl1_def}) into a series about $y^n=0$ and then to calculate the integral from $0$ to $1$. 
Inserting $1=(1-u) + u$ and integrating by parts, we obtain the first term of the asymptotic series about $y^n = 0$:
\begin{equation}
T_{l,1}^{n,m}(y^n) = \frac{1+O(y^n)}{n} \left\{ \begin{array}{lr}
- \frac{y^{n\delta}}{\delta} & {\rm if \ } \delta<0 \\
-\ln y^n & {\rm if \ } \delta=0 \\
(1-2\delta) B(\delta,\frac{1}{2})  + 2\beta  B(\beta,\frac{1}{2})  & {\rm if \ } \delta>0 \\
\end{array} \right.
\label{tl1}
\end{equation}
 with 
$$\delta= \beta + \frac{n-m}{n} = \frac{3n - 2m -2 l + 2}{2n}$$

Let us focus hereafter on  the first order correction in $\alpha_c$. Equation~(\ref{eqn_y0}) leads to
$
y^n \approx y_0^n (1- \alpha_0)
$
and the formula~(\ref{equ_fin_u}) becomes: 
\begin{equation}
 I_l^{\rm a}  =  (-C_n)^{ -1/2 }  \left( \frac{(R_+^c)^n}{y_0^n} \right)^{\beta} 
\left[ 
 T_{l,0}^{n,m}(y_0^n(1- \alpha_0)) (1+\beta \alpha_0)   - \frac{1}{2} \alpha_0 T_{l,1}^{n,m}(y_0^n)
\right]
\label{final_formula}
\end{equation}
where we have to consider only terms  in formulas~(\ref{tl0}) and (\ref{tl1})
 without the $O(y^n)$ term. We have  neglected terms that would be part of the correction in second order in $\alpha_c$, and  would lead to an explicit dependence of the NDE formulas on the cut-off $R_+^c$ (which then won't be
 no longer hidden in the $\tilde \gamma_\delta$ term). But, this might not be the best strategy. We know that $y=R_+^c/R_+$ is not necessary small (but $y^n$ will be). It might be better to approximate the integrals (\ref{tl1_def}) by Pade  series in $y^n$ than by Taylor series. Another method would be to obtain  an analytical solution for the integral (\ref{int_tot}),  using for instance a third term (as for instance $\frac{-C_m^2}{ C_n R^{2m-n}}$) in the multipole expansion (\ref{Dv_mul_cut_off}).

Finally, with those choices, the full (non asymptotic) formula reads:
\begin{eqnarray}
 I_l  &\approx& (-C_n)^{ -1/2 }\frac{ B(\beta,1/2)}{n} \left(  \frac{D-E}{-C_n} \right)^{-\beta}
+ \left( I_l^\na (D) -  (-C_n)^{ -1/2 }\frac{(R_+ ^c)^{n \beta } }{n\beta} \right) + \label{final_formula_ann}
 \\
& &  \frac{(-C_n)^{ -1/2 }}{n} \frac{C_m}{C_n}  \left(  \frac{D-E}{-C_n} \right)^{-\delta} \Big[ \beta B(\beta, \frac{1}{2} )+  \left\{ \begin{array}{lr}
\frac{1}{2\delta}    \left(  \frac{D-E}{-C_n/(R_+^c)^n} \right)^{\delta} & {\rm if \ } \delta<0 \\
\frac{1}{2} \ln (D-E) + \frac{1}{2}   \ln \frac{(R_+^c)^n}{-C_n}  & {\rm if \ } \delta=0 \\
( \delta - \frac{1}{2} )  B(\delta,\frac{1}{2}) - \beta  B(\beta,\frac{1}{2}) & {\rm if \ } \delta>0 \\
\end{array} \right.
\Big] \nonumber  
\end{eqnarray}
where to summarize our calculation of formula (\ref{int_tota}), we have  used the assumption $E\approx D$ in the non-asymptotic
part $\int_{R_-}^{R_+^c}$.

\section{Long-range states.}
\label{long_range}

We shall give a brief introduction to the Hund's  case (c) states and potential curves  calculation for long-range molecules. We shall
mainly discuss the case of neutral alkali-atoms. The discussion can easily be extended to all atoms. We refer to the following books and articles~\cite{Marinescu95,Marinescu96,Herzberg,Hougen,Nikitin,LFB,Marinescu99,cm:aubertfrecon98,cm:aubertfrecon98b}.

\subsection{Non relativistic electrostatic interaction and multipolar development}
In the following,  we shall discuss the interaction between two atoms $A$ and $B$ (radius $r_A$ and $r_B$) formed by a core $A^+$ and 
a valence electron $e_1$ for the first one and by a core $B^+$ and a valence electron $e_2$ for the second one.

The first assumption for the Born-Oppenheimer Hund's  case (a) potential curve calculation is to neglect all the relativistic parts. Then the atomic Hamiltonian $H^{\rm at}_A$ for atom $A$  leads to  $| 1: n l m_l
m_s \rangle_A$ states (quantize on the internuclear axis) represented by the eigenfunction: 
\begin{equation}
\Psi^A_{n l m_l} (\vec r_{A 1} ) \chi_{s=1/2,
m_s} =  { u^A_{n l} (r_{A 1} ) \over r_{A 1} } 
 Y_l^m ( \hat r_{A 1} )
\chi_{s=1/2, m_s}
\label{rep_stat}
\end{equation}
 where $\hat r_{A1} $ are the polar angle of $\vec r_{A1}$.
Then the eigensystem of the full electrostatic Hamiltonian 
\begin{equation}
H=H^{\rm at}_A + H^{\rm at}_B + H^{\rm el}
\label{full_ham}
\end{equation}
can be computed. We will use the first and second order perturbation theory for $H^{\rm el}$.

The large internuclear distance assumption ($R >> r_A + r_B$) leads to a Taylor series about $1/R$. It is convenient to define the $2^j$ polar irreducible tensors $Q^j$ by (for atom $A$):
\begin{equation}
Q_j^m = \sqrt{ {4 \pi \over 2j+1 }} 
q_e r_{A1}^j  Y_j^m ( \hat r_{A1} )
\label{def_irre_tens}
\end{equation}
For a numerical computation of the matrix element, this expression should be multiply by $1+ \epsilon(r_{A1} )$ in order to
 take into account the effect of all electrons and not only the valence one~\cite{Marinescu94}.

With these notations, the Taylor series is~\cite{Dalgarno66,Bussery85}:
$
H^{\rm el}  = 
\sum_{i,j=0}^\infty { V_{ij} (A,B) \over R^{i+j+1} }
$
where
\begin{eqnarray}
\lefteqn{ V_{ij} (A,B)  =  {1 \over 4 \pi \varepsilon_0} \times } \label{Ham_mol_mul_WE}  \\
& &
\sum_{m=- \inf(i,j)}^{ \inf(i,j)} { (-1)^j (i+j) ! Q_i^m (A) Q_j^{-m} (B)
 \over \sqrt{
(i+m) ! (i-m) ! (j+m) ! (j-m) ! } } \nonumber 
\end{eqnarray}
It is then straightforward to see, for instance, 
dipole-dipole interaction has a $R^{-3}$ behavior.
 
\subsection{Molecular symmetries and Hund's  case (a) states}

It is useful to use the molecular symmetries for $H^{\rm el}$. Because for us the nuclear spin is passive (see~\cite{Comparat00,BoGao} for details), we will not study the symmetries for
the total (electronic, rotation, vibration, nuclear spin) wavefunctions, as the permutation for bosons or fermions nucleus, but only for the electronical part:

\begin{itemize}
\item 
Orbital electronical rotation around the molecular axis $Oz$, leading to a well defined projection $\hbar m_L= \hbar \Lambda$ on $Oz$.

\item
Orbital electronical reflexion $\sigma'_v$  versus $yOz$ (or a different plan~\cite{LFB}) (eigenvalue $\sigma' = \pm 1$). This commutes with the previous one if $\Lambda =0$.

\item
In the case of identical nuclear charge $Z_A=Z_B$: electronical orbital inversion $
I^{\rm el}  f(\vec r_{A 1},\vec
r_{B 2}) = f(- \vec r_{B
1},- \vec r_{A 2})
$ with eigenvalues $\omega'=\pm 1$ and states noted $g$ ({\em gerade}) for $\omega'=1$  and $u$ ({\em ungerade}) for $\omega'=-1$.

\item 
Rotation of the electronical spin $\vec S$. So ${\vec S}^2$ and $S_z$ are eigen operators with $\hbar^2 S(S+1)$ and $\hbar m_S= \hbar \Sigma$ eigenvalues.

The electronical spin is passive so it is easier to separate it and to define molecular spin states $|S, m_S=\Sigma\rangle$  by:
$$
\begin{array}{lcl} |0,0\rangle & = &
{1 \over \sqrt{2} } \big( |+ - \rangle - |- + \rangle \big)
\label{base_adap} \\
|1,-1\rangle & = & |- - \rangle \\ |1 , 0 \rangle & = & {1 \over
\sqrt{2}} \big( |+ - \rangle + |- + \rangle \big) \\ |1,1\rangle
& = & |+ + \rangle \\ \end{array}
$$
with the convention $m_{s_{e_1}}$ is noted in first place in the $| m_s m'_s \rangle$ notation.

\end{itemize}

 Spherical harmonics formulas~\cite{Var} for the atomic wavefunction~(\ref{rep_stat}) lead to:
\begin{eqnarray}
\lefteqn{ \sigma'_v | 1: n l m_l m_s\rangle_A 
 |2: n' l' m'_l m'_s\rangle_B  = } \label{reflex_for} \\
& &  | 1: n l -m_l m_s\rangle_A \otimes |2: n' l' -m'_l
m'_s\rangle_B \nonumber \\
\lefteqn{
I^{\rm el} | 1: n l m_l m_s\rangle_A  |2: n' l'
m'_l m'_s\rangle_B = } \\
& & (-1)^l (-1)^{l'} | 2: n' l' m'_l
m'_s\rangle_A  |1: n l m_l m_s\rangle_B
\end{eqnarray}

 Electrons are fermions so the final state has to be anti-symmetrical
for the electronical exchange $P_{1 2} = P_{1 2}^{\rm orb} P_{1 2}^{\rm spin} $:
\begin{eqnarray*}
\lefteqn{
 P_{1 2} |1
: n l m_l m_s\rangle_A \otimes |2: n' l' m'_l m'_s\rangle_B = }\\
& &
|2: n l m_l m_s\rangle_A \otimes |1: n' l' m'_l m'_s\rangle_B
\end{eqnarray*}

Finally, with the usual Hund's  case (a) notations, the molecular state we want to calculate is:
$$
\left|
^{2S+1}
 \Lambda_{\Omega,(\omega')}^{(\sigma')}  \right\rangle$$
where $\Omega=\Lambda+\Sigma$. 
Some other basis, as the Wang  basis~\cite{LFB} with $|\Lambda|$, can also be used, as all calculations can be done in any complete basis ; the one we choose  leads to simple expressions.

We are focusing on this paper to the states reached by photoassociation. Most of the photoassociation experiments  start with two atoms in ground state 
($ns + ns$ state)   photoassociated toward the first excited asymptotes $ns +np$. Therefore, we shall continue the discussion with one atom ($A$ or $B$) in $n l$ state and the other one in $n' l'$ state with $l$ or $l'$ null (but $n\not = n'$ is not required). 
It is nevertheless simple to obtain the  formulas for the general $nl+n' l'$ configuration~\cite{Nikitin,BoGao,Zygelman}.

From the previous expression for symmetry operators,  it is easy to verify that the wanted expression is:
\begin{eqnarray}
\lefteqn{ \left|
^{2S+1}
 \Lambda_{\Omega,\omega'}^{(+)}  \ (ns +n' l') \right\rangle^0 = }
\label{etat(a)6s+6p}  
\\ & &   \bigg[ { 1 +
(-1)^S P_{1 2}^{\rm orb} \over \sqrt{2} }  c \big ( | 1: n 0 0
\rangle_A |2: n' l' \Lambda \rangle_B  + \label{hund_a} \\
& &
\omega' (-1)^{-S-l'} | 1: n' l'
\Lambda \rangle_A |2: n 0 0 \rangle_B \big) \bigg]
|S,\Omega- \Lambda \rangle 
\nonumber
\end{eqnarray}
where the $^0$ exponent means we did not use
 the perturbation theory yet,  so this state is the zero order state for a given internuclear distance $R$.
$c$ is a normalization constant slightly $R$-dependent due to the exponential overlap between $\Psi^A (\vec r + \vec R )$ and $\Psi^B (\vec r)$. Similarly , the $P_{1 2}^{\rm orb}$ operator will put the electron $e_1$ (resp. $e_2$) close to the core $B^+$ (resp. $A^+$): this leads to another exponential correction, known as the exchange correction~\cite{Marinescu96,Hadinger,Aubert99}.

We will work with large enough internuclear distances to avoid the overlap and exchange terms. We can then make use of ${ 1 +
(-1)^S P_{1 2}^{\rm orb} \over \sqrt{2} }  =  1$  in expression~(\ref{hund_a}) for the multipole coefficient calculation and $c=1/\sqrt{2}$ when $(n,l) \not =( n', l' )$, $c=1/2$ otherwise. 
Several authors tried to estimated what is the right cut-off to safely neglect the overlap and exchange effects. Let us
 just mentioned the article~\cite{Ji} modifying the simple 1973 LeRoy  limit:
$$
R > 2 (r_A + r_B)
$$
leading to typical values close to $20\,a_0$.

Among the 24 $ns+n'p$ states only 16 will be non-degenerate as shown in figure~\ref{fig:hundc}.

 \subsection{Interactions, exchange and Hund's  case (c) curve}
\subsubsection{Interactions and multipole coefficients for Hund's  case (a) curves}

The next step, starting with this zero order basis, is to apply the perturbation theory, or better, the degenerate perturbation theory (see~\cite{Marinescu95,Marinescu96}),
 to the $H^{\rm el}$  perturbation.

Calculation is straightforward using formula~(\ref{Ham_mol_mul_WE}).
As before  we detail only formulas
 for the $ns+n'p$ asymptote. This choice yields for the first order perturbation to  $i=j=1$, i.e. a dipole-dipole interaction, and  energy:
\begin{eqnarray}
\lefteqn{ V_{^{2S+1} \Lambda_{\Omega,\omega'}^{(+)} }  } \\
& = &
 ^{0} \left\langle ^{2S+1} \Lambda_{\Omega,\omega'}^{(+)} ns+n'p \right|
H^{\rm el}
 \left| ^{2S+1} \Lambda_{\Omega,\omega'}^{(+)}  ns+n'p \right\rangle^0 \nonumber \\
 & & = - \omega'(-1)^{-S-1} {2 (-1)^\Lambda \over 1 + |\Lambda| } {C_3 \over R^3} \nonumber
\end{eqnarray}
where 
\begin{eqnarray}
C_3 & =& \left( { D^2 \over 12 \pi
\varepsilon_0 } \right) \label{C3coef}, \hbox{ and the dipole is} \\
D & = & \langle ns \parallel Q_1
\parallel n'p \rangle
= - \langle n'p \parallel Q_1
\parallel ns \rangle \nonumber \\
& = &  \sqrt{3} \langle n'p m_l=0 | q_e z | n s m_l=0 \rangle
\nonumber
\end{eqnarray}
Second order perturbation theory  leads to the so called polarization terms (London or dispersion, and Debye or induction) yielding  the
 final multipole expansion formula:
\begin{equation}
V_{^{2S+1} \Lambda_{\Omega,\omega'}^{(+)} } (R) =  E_p +
{C'_3 \over R^3}+ {C'_6 \over R^6} + {C'_8 \over R^8} + \ldots
\label{dev_mul_c}
\end{equation}
where $E_p $ is the dissociation limit energy $ns + n'p$.
Theoretical value for several $C'_6$ and $C'_8$  coefficients can be found in~\cite{Marinescu95}.

\subsubsection{Hund's  case (c) versus Hund's  case (a) states}
For heavy atoms we must add another perturbation term, the spin-orbit one, in the Hamiltonian~\cite{LFB} 
$$
H^{\rm SO} =
A^{\rm SO} {l_1^+ s_1^- + l_1^- s_1^+ \over 2} + {l_1}_z {s_1}_z
+ 1 \leftrightarrow 2 
$$
where $A^{\rm SO}$ is the atomic spin-orbit constant. Due to
spin-other orbit ($e_2$ on $A^+$ and $e_1$ on $B^+$) or mixing with curves coming from $n'' l''+n'''l'''$ dissociation limits, 
$A^{\rm SO}$ is in fact slightly $R$ dependent.

The new Hamiltonian results in less symmetries. Only $\vec J = \vec L + \vec S$ is a good operator leading to $\Omega$ as a good quantum number. Furthermore the
 electronical reflexion $\sigma_v$ has to act on spin also. It is straightforward~\cite{Nikitin,Messia} to see its eigenvalues verify $\sigma =  (-1)^S \sigma'$.
It is then better to work with a new basis:
\begin{eqnarray*}
\left| ^{2S+1} |\Lambda|_{|\Omega|,\omega'}^\sigma \right\rangle^0 & = &{1 \over \sqrt{2} } (
\left| ^{2S+1} \Lambda_{\Omega,\omega'}^{(+)} \right\rangle^0 + \\
& & (-1)^S \sigma
\left| ^{2S+1} -\Lambda_{-\Omega,\omega'}^{(+)} \right\rangle^0)
\end{eqnarray*}
This definition is valid  for $\{ \Lambda,\Omega \} \not= \{ 0 , 0 )\}$, on the contrary
$\left| ^{2S+1} \Lambda_{\Omega,\omega'}^{(+)} \right\rangle^0$ is already an eigenstate for $\sigma=(-1)^S$.

 Using $H^{\rm SO}$ definition and formula~(\ref{hund_a}) yields the block matrix
$1\times 1$ for $|\Omega| = 2$,
$2\times 2$ for $|\Omega| = 0$ and
$3\times 3$ for $|\Omega| = 1$.
Finally, we only have to diagonalize these matrices, given in table~\ref{table_etat(c)}, 
to have Hund's  case (c) states $\Omega_{\omega'}^\sigma$ and potential curve versus the known Hund's  case (a) ones.

\begin{table}
\caption{Matrix given
$\Omega_\gu^{ \pm }$ Hund's  case (c) eigensystem ($R$ variation is not written) for $ns+
n'p$ limit. With $A=\hbar^2 A^{\rm SO}  = (E_{p_{3/2}} - E_{p_{1/2}} ) 2/3$.} 
\begin{eqnarray}
 \pm 2_\gu   &= &  V( ^3 \Pi_{\pm 2,\gu} )
+ { A \over 2}  \nonumber \\
 & & \nonumber \\
\pm 1_\gu &=& \left( \begin{array}{ccc} V ( ^3  \Pi_{ \pm 1,\gu} )
& {\mp A \over 2 } & { A \over
2 } \\ {\mp A \over 2 } & V ( ^1 \Pi_{\pm 1,\gu} )
& {\pm A \over 2 }\\
 { A \over
2 } & { \pm A \over 2 } & V ( ^3 \Sigma_{\pm
1,\gu}^+ ) \end{array} \right) \nonumber \\
0_\gu^+ & = & \left( \begin{array}{cc} V (^3 \Pi_{0,\gu}^{\sigma =
+} ) - {A \over 2} &  {- A
\over \sqrt{2}} \\
 {- A \over  \sqrt{2}}  & V (
^1 \Sigma_\gu^+ ) \end{array} \right)  \nonumber  \\
0_\gu^- &=& \left( \begin{array}{cc} V (^3 \Pi_{0,\gu}^{\sigma =
-} ) - {A \over 2} & {A \over 
\sqrt{2} } \\ {A \over \sqrt{2}}  & V ( ^3
\Sigma_{0,\gu}^+ ) \end{array} \right) 
\nonumber
\end{eqnarray}
\label{table_etat(c)} 
\label{tabl}
\end{table}

\subsubsection{Hund's  case (c) versus Hund's  case (e) states}

In heavy alkali atoms, $H^{\rm SO}$ is quite larger than
$H^{\rm el}$  at large internuclear distance. Consequently it is better to work with a "fine structure" Hund's  case (e) basis $|1
: n l j m_j\rangle_A \otimes |2: n' l' j m'_j\rangle_B$ where  $H_A^{\rm at} + H_B^{\rm at} + H^{\rm SO}$ is diagonal.
 
To calculate the only missing perturbation $H^{\rm el}$,  we need, as before, to find a well (molecular)-symmetrized   basis~\cite{Marinescu96,Nikitin}. To avoid such a work, we take advantage on the already known matrix given in table~\ref{table_etat(c)}.
By definition the Hund's  case (e) basis is just formed by the eigenstates of the matrices given in table~\ref{table_etat(c)} when only the spin-orbit is present (all the electrostatic interactions  $V_\Pi,V_\Sigma$ are set to zero).

Let us just illustrate it on the $0_g^-$ case. The transition matrix is:
$$P_{a\rightarrow e}=
\left(  \begin{array}{cc} 
 {1 \over \sqrt{3}} & \sqrt{2 \over 3} \\
 \sqrt{2 \over 3} & - {1 \over \sqrt{3} } 
 \end{array} 
\right)
$$
Then the desired matrix of $H^{\rm el}$ in the Hund's  case (e) basis is just
$
M_{(e)}^{0_g^-}   =  P M_{(a),H^{\rm SO}+H^{\rm el}}^{0_g^-}  P^{-1} $.
 It is then better to use reduce matrix element related to $n'p_j$ \cite{Var}:
\begin{eqnarray}
D_{3/2}&=&\langle ns_{1/2} \parallel Q_1 \parallel n'p_{3/2} \rangle  = 
{2 \over \sqrt{3} }
\langle ns \parallel Q_1 \parallel n'p\rangle \label{D32}
\\ 
D_{1/2}&=&\langle ns_{1/2} \parallel Q_1 \parallel n'p_{1/2} \rangle =
 {\sqrt{2} \over \sqrt{3} }
\langle ns  \parallel Q_1 \parallel n'p\rangle
\label{D12}
\end{eqnarray}
 to have the new matrix:
\begin{eqnarray}
M_{(e)}^{0_g^-} 
& = & \left(  \begin{array}{cc} E_{p_{3/2}} - {1 \over 16 \pi \varepsilon_0} 
{D_{3/2}^2 
\over R^3 }
   & 
{1  \over 8 \pi \varepsilon_0}
 
{  D_{3/2} D_{1/2} \over  R^3 }
\\
{1 \over 8 \pi \varepsilon_0} 
{  D_{3/2} D_{1/2} \over R^3 }
 & 
E_{p_{1/2}}
 \end{array} 
\right) \label{mat_cas_e_relat}
\end{eqnarray}
where, for sake of simplicity, we have kept only the $C'_3/R^3$ leading term in expression~(\ref{dev_mul_c}).
As expected the spin-orbit is diagonal leading to the dissociation limit toward
$ns + n'p_{3/2}$ with an energy $E_{p_{3/2}} = E_p + { \hbar^2 A^{\rm SO} \over 2}$
and toward $ns + n'p_{\rm 1/2}$ with an energy $E_{p_{1/2}} = E_p - \hbar^2 A^{\rm SO} $. 

\subsubsection{Relativistic correction}
 This new matrix is also useful, as we are going to see, to add a relativistic correction to the $C_3$ coefficient in the case $n=n'$.

Indeed, the experimental lifetime measurements noted $\tau_{3/2}$ and $\tau_{1/2}$ for $np_{3/2}$ and $np_{1/2}$~\cite{Rafac99,Volz,Rafac}
  disagree with the predicted ratio  $D_{3/2}/D_{1/2} = \sqrt{2}$  (see formulas~(\ref{D32}) and~(\ref{D12})). This relativistic
correction has to be taken into account. Let us define the $\epsilon$ correction by
\begin{equation}
{
D_{1/2} 
\over
D_{3/2} 
} = (1 + \epsilon)  \sqrt{ 1 \over 2} = \sqrt{ {\tau_{3/2} E_{3/2}^3 \over 2 \tau_{1/2} E_{1/2}^2 } } \approx \sqrt{ 1 \over 1.9809(9) }
\label{def_epsilon}
\end{equation}
where the experimental value is given for cesium.
A new definition for $C_3$:
\begin{equation}
C_3 =
{1 \over 4}
{1 \over 4 \pi \varepsilon_0}  | \langle ns_{1/2} \parallel Q_1 \parallel np_{3/2} \rangle |^2 
\label{new_c_3_def}
\end{equation}
leads to the same value than the previous one (see formula~(\ref{C3coef})) for $\epsilon =0$  (the value for cesium is $C_3 =9.997(23)$~\cite{Rafac99}).

Then the new matrix~(\ref{mat_cas_e_relat}) is:
$$ 
M_{(e) }^{0_g^-} 
 =  \left(  \begin{array}{cc}  E_{p_{3/2}}
 - {C_3 \over R^3}
   & 
\sqrt{2} {C_3 (1 + \epsilon) \over R^3} \\
\sqrt{2} {C_3 (1 + \epsilon) \over R^3}
 & 
E_{p_{1/2}}
 \end{array} 
\right) 
$$
Using the transition matrix $P$ we can then include the perturbation in the Hund's  case (a) matrix to find:
\begin{equation}
M_{(a)}^{0_g^-}  = \left(  \begin{array}{cc} E_p+ {C_3 (1+ {4 \epsilon \over 3} )  \over 
R^3} - { \hbar^2 A^{\rm SO} \over 2}   & 
{ \sqrt{2} \over 3} \epsilon {C_3  \over 
R^3} +
{\hbar^2 A^{\rm
SO} \over  \sqrt{2} } \\ 
{ \sqrt{2} \over 3} \epsilon {C_3  \over 
R^3} + 
{\hbar^2  A^{\rm SO} \over
\sqrt{2}  } & E_p -2 {C_3  (1+ {2 \epsilon \over 3} )\over R^3}
 \end{array} 
\right)
\label{matriceog-} 
\end{equation}
Comparing this matrix to the one calculated without the $\epsilon$ correction 
 leads to the correction matrix $\delta M_{(a)}$  given in table~\ref{corr_matrix}.
In the cesium case $\epsilon\approx 0.005$ therefore the corrections 
factor $(1+\epsilon)$ (or even more  $(1+\epsilon)^2$) are (just) needed to have the percent accuracy we are dealing with.

Their is a second relativistic correction known as the  retardation effect~\cite{Stephen,Meath,Gomberoff} (see also reviews~\cite{SpruchCas,Power67} and recent articles~\cite{MDoct94,MY99}). The main retardation correction for $\Sigma$ (respectively $\Pi$) states concern the $C_3$ coefficient which has to be multiply by $f_\Sigma$ (respectively $f_\Pi$) where:
\begin{eqnarray*}
f_\Sigma &=&
\cos {R
\over \bar \lambda } 
 +  {R  \over \bar \lambda } \sin {R  \over \bar \lambda } \approx 1 + {1
\over 2} {R^2 \over \bar \lambda^2} \\
 f_\Pi &=&  -{R^2 \over
\bar \lambda^2} \cos { R \over \bar \lambda}   +{ R \over \bar
\lambda} \sin { R \over \bar \lambda} + \cos { R \over \bar
\lambda} \approx 1 - {1 \over 2} {R^2 \over \bar \lambda^2}
\end{eqnarray*} 
where $\bar \lambda = c/\omega $ ($2400\,a_0$ for Cs) and $ \hbar \omega=E_p-E_s$. This theory is limited for  $R<c \tau$ (several centimeters) due to the photon lifetime.

We will consider this correction as almost negligible for our purpose of $1\%$ accuracy because our description will focus on $R<200\,a_0 \approx \bar \lambda/12$.

\begin{table}
\caption{The relativistic correction $\delta M$ to the matrix proposed in table~(\ref{tabl}); 
$\epsilon$ value is  defined by formula (\ref{def_epsilon}).  States $g$ correspond to $\omega'=1$  and $u$ to $\omega'=-1$.}
\begin{eqnarray} 
\delta \pm 2_\gu  &=& 0 
\nonumber \\ 
\delta \pm 1_\gu &=&    (-1)^{\omega'+1} \frac{\epsilon C_3 }{9 R^3} 
\left( \matrix{
  2(\epsilon -3 ) & \pm 2 \epsilon & -( 3 + 2\epsilon  ) \cr
 \pm 2\epsilon & 2( 3 + \epsilon  )  & \mp ( 9 + 2\epsilon ) \cr
  -( 3 + 2\epsilon ) & \mp ( 9 + 2\epsilon) & 2( 6 + \epsilon )
\cr  } \right)
 \nonumber  \\
\delta 0_\gu^+ &=& (-1)^{\omega'+1} \frac{\epsilon C_3 }{9 R^3}  
\left( \matrix{
-4( 3 + 2\epsilon  )& - \sqrt{2}( 9 + 4\epsilon ) \cr 
-\sqrt{2}( 9 + 4\epsilon ) & -4( 3 + \epsilon  )\cr  } \right)
 \nonumber \\
\delta 0_\gu^- &=& (-1)^{\omega'+1} \frac{\epsilon C_3 }{3 R^3}  \left( \matrix{ 
-4 & -\sqrt{2} \cr 
-\sqrt{2} & 4 \cr  } \right)
\nonumber
\end{eqnarray}
\label{corr_matrix}\end{table}

To conclude this calculation, and in order to obtain a precise potential curve determination, we have to include some other small effects. These effects are usually negligible
to obtain the two leading terms in the multipolar extension as it is needed in our NDE expressions. Therefore they should not have any incidence in our calculation.

\begin{enumerate}
\item Spin-orbite $R$ (dependence), exchange, overlap.

These effects will mainly affect the intermediate part of the potential curve and not the "pure-"long range part we are interested for our asymptotic calculation.  
\item Spin-spin.

As previously discussed, the 
spin relativistic effect has to be taken into-account for a precise potential curve determination~\cite{Gomberoff}. Spin-rotation, dipole-(spin-dipole) are negligible. The spin-spin interaction leads for instance for the $0_g^-$ matrix to the correction:
\begin{equation}
 {\hbar^2 e^2 \over m_e^2 c^2 R^3}  \left( 
 \begin{array}{cc} 
-{1 \over 2}  & 0 \\
  0 &  1
 \end{array} 
\right) \label{mat_spin_spin}
\end{equation}
This is a negligible term in the multipolar development because, for cesium,
${\hbar^2 e^2 \over m_e^2 c^2} \approx C_3\times 5.10^{-6}$.

\item Rotation and Coriolis.
The rotational part is given by 
$$
H^{\rm rot} = { \ell^2 \over 2 \mu R^2} = { (\vec J - \vec L - \vec S)^2 \over 2 \mu R^2}
$$
This usual derivation~\cite{LFB} yields for instance in the $0_g^-$ case  the matrix correction:
$$ 
 {\hbar^2 \over 2 \mu R^2 }  \left( 
 \begin{array}{cc} 
J (J+1) + 2 & 2 \sqrt{2} \\
2 \sqrt{2} & 4 + J (J+1) \end{array} 
\right)
$$

For $R \approx 200\,a_0$ we have  ${\hbar^2  \over 2 \mu R^2 } \approx 10^{-4} {C_3 \over R^3}$. This correction is also negligible for the multipole expansion.

\item Kinetics coupling and mass polarization terms.

These terms lead typically to a correction of less than one percent~\cite{MDmarch98}.
\end{enumerate}

\end{document}